\documentclass[traditabstract]{aa}
\usepackage{natbib}
\usepackage{txfonts}
\usepackage{graphicx}

\begin{document}

\title{First evidence of a magnetic field on Vega}
\subtitle{Towards a new class of magnetic A-type stars\thanks{Based on observations at Telescope Bernard Lyot of Observatoire du Pic du Midi, CNRS/INSU and Universit\'e de Toulouse, France}}	

\author{F. Ligni\`eres \inst{1,2} \and P. Petit \inst{1,2} \and T. B\"ohm \inst{1,2} \and M. Auri\`ere \inst{1,2}}
\institute{
Universit\'e de Toulouse; UPS; Laboratoire d'Astrophysique de Toulouse-Tarbes (LATT); F-31400 Toulouse, France 
\and
CNRS; Laboratoire d'Astrophysique de Toulouse-Tarbes (LATT); F-31400 Toulouse, France}

\date{Received March 6, 2009/ Accepted April 29, 2009}

\abstract
{}
{We report the detection of a magnetic field on Vega through spectropolarimetric observations.}
{We acquired 257 Stokes V, high signal-to-noise and high-resolution echelle spectra during four consecutive nights with the
NARVAL spectropolarimeter at the 2-m Telescope Bernard Lyot of Observatoire du Pic du Midi (France).
A circularly polarized signal in line profiles is unambiguously detected after combining the contribution of about $1200$ spectral lines for each
spectrum and summing the signal over the 257 spectra.
Due to the low amplitude of the polarized signal, various tests
have been
performed to discard the possibility of a spurious polarized signal. They all point towards a stellar origin of the polarized signal.}
{Interpreting this polarization as a Zeeman signature leads to a value of $-0.6 \pm 0.3$~G for the disk-averaged line-of-sight
component of the surface magnetic field.
This is the first strong evidence of a magnetic field in an A-type star which is not an Ap chemically peculiar star.
Moreover, this longitudinal magnetic field is smaller by about two
orders of magnitude than the longitudinal magnetic field (taken at its maximum phase) of the most weakly magnetic Ap stars.
Magnetic fields similar to the Vega magnetic field could be present but still undetected in many other A-type stars.
}
{}

\keywords{stars: magnetic fields - stars: early-type, stars: individual: Vega}

\maketitle

\section{Introduction}

Despite recent progress in stellar magnetic field measurements, 
spectropolarimetric surveys of
early-type stars indicate that 
photospheric magnetic fields can only be detected in a small fraction of these stars.
Without direct constraints on the magnetic field of the vast majority of early-type stars, our understanding
of the role of magnetic fields on the structure and
evolution of intermediate mass and massive stars is necessarily limited.
In this Letter, we report the detection of a magnetic field on Vega and argue that Vega is probably the first member of 
a new class of yet undetected magnetic A-type stars.

The proportion of stars hosting a detectable magnetic field is more firmly established for main sequence stars of intermediate mass 
(late-B and A-type stars) than for massive stars (early B and O-type stars) or intermediate mass pre-main-sequence stars (Herbig Ae/Be stars). 
Magnetic A-type stars are indeed identified with the group of Ap-Bp chemically peculiar stars 
(excluding the subgroup of HgMn stars) since all known magnetic A-type stars belong to this group and, when observed with sufficient precision, 
Ap/Bp stars always show photospheric magnetic fields \citep{Lan92,Au07}.
The incidence of the Ap/Bp chemical peculiarity among A-type stars then leads to
a $5-10\%$ estimate of magnetic stars \citep{Wo68}.
Note that magnetic field detections have been reported for a few Am and HgMn stars \citep{Ma93,Ma95} but remain debated because
they could not
be confirmed by further investigations
\citep[see the discussion in][]{Shor02}.
Thanks to new high-resolution spectropolarimeters, magnetic fields are now also detected in pre-main-sequence stars
and in massive stars.
According to recent surveys, the fraction of magnetic stars among Herbig Ae/Be stars is $7 \%$ \citep{Wade09}, 
while the rate of detection for early B and O-type stars is also small 
\citep{Bou08,Schnerr08}.

The magnetic fields of Ap/Bp stars are characterized by a strong dipolar component, a long-term stability
and dipolar strengths ranging from a lower limit of about 300 Gauss to tens of kilo-Gauss \citep{Lan92,Au07}.
Thus, if a population of weak dipolar-like fields corresponding to a
weak field continuation of Ap/Bp stars exists, a longitudinal component of the magnetic field in the range of $10$ to $100$ Gauss
should have been detected by recent spectropolarimetric surveys of non Ap/Bp stars \citep{Shor02,Wade06,Ba06,Au09}.
Instead, these surveys suggest there is a dichotomy between the population of strong, stable and dipolar-like
magnetic fields corresponding to the Ap/Bp stars and the rest of A-type stars, whose magnetic properties remain unknown, except
that their surface longitudinal magnetic field should be very small.

Vega is well suited for the search of magnetic fields among A-type non Ap/Bp stars.
Its brightness and its low equatorial projected velocity 
ensure high signal-to-noise V spectra, while the number of spectral lines of an A0-type star is important enough to
allow a very large multiplex
gain by gathering the polarimetric signal of
all the lines using a cross-correlation technique (Least-Squares Deconvolution, \citet{Do97},
LSD hereafter).
Another advantage of Vega's brightness is that its fundamental parameters are well known relative to 
other more anonymous stars \citep{Gr07}.
In particular, spectral analysis and interferometric observations have shown
that Vega is a rapidly rotating star seen nearly pole-on \citep{Auf06,Pe06,Ta08}.

Vega was already included in a previous spectropolarimetric survey of A-type non Ap/Bp stars using NARVAL at the Telescope Bernard Lyot 
of Pic du Midi, but the analysis of its 11
Stokes V spectra was not conclusive. Here we present the results of 
a four night observing run fully dedicated to Vega, during which more than 300 Stokes V spectra were obtained.
Summing the information over a large number of these spectra leads to an 
unambiguous detection of a polarized signal.

The observations are described and interpreted in the next section. In section 3, the 
origin of
a $\sim \!1$~G longitudinal magnetic field in an A-type non Ap/Bp star is discussed and some of the perspectives opened by this field detection are considered.
Our conclusions are given in section 4.

\section{Instrumental setup, data reduction and multi-line extraction of Zeeman signatures}

The observing material was gathered at the Telescope Bernard Lyot (Observatoire du Pic du Midi, France) using the NARVAL spectropolarimeter. 
As a strict copy of ESPaDOnS \citep{Pet03}, NARVAL spectra provide a simultaneous coverage of the whole optical domain (from 370 nm to 1,000 nm) 
at high spectral resolution ($R=$65,000). The instrument consists of a bench-mounted spectrograph and a Cassegrain-mounted polarimeter,
with an optical fiber carrying the light between the two units. A series of 3 Fresnel rhombs (two half-wave rhombs that 
can rotate about the optical axis and one fixed quarter-wave rhomb) are used in the polarimeter, followed by a Wollaston prism which splits the 
incident light into two beams, respectively containing light linearly polarized perpendicular/parallel to the axis of the prism. 
Each Stokes V spectrum is obtained
from a combination of four sub-exposures taken with the half-wave rhombs oriented at different azimuths \citep{Sem93}. 
The data reduction is performed by Libre-Esprit, a dedicated, fully automated software described by \citet{Do97}.

The data were collected during 4 consecutive nights in the summer of 2008, from July 25 to July 28, using 6~sec integration times for each
sub-exposure of the Stokes V sequences (except the first two sequences of the run, for which exposure times of 15 and 10 sec were adopted). We 
retained the
257 Stokes V spectra with a typical peak signal-to-noise ratio (S/N hereafter) of 1,500 
per 1.8 ~km\,s$^{-1}$, around $\lambda = 600$~nm.

\begin{figure}
\centering
\includegraphics[width=10cm]{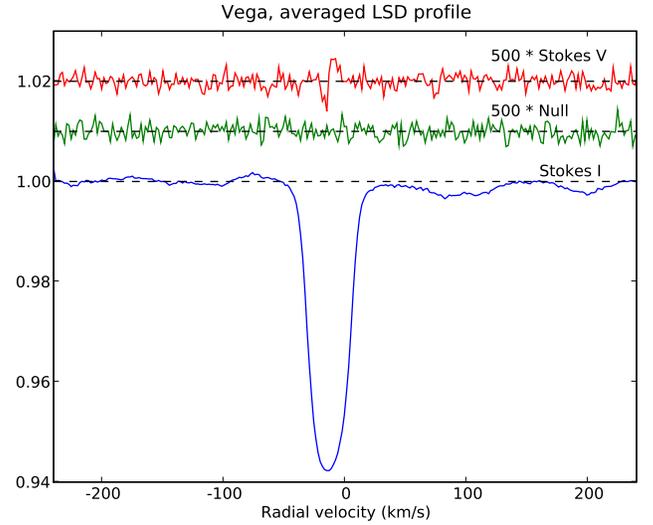}
\caption{(Color online) Average of the 257 normalized stokes I (blue/bottom) and Stokes V (red/upper) LSD profiles of Vega, as a function of the radial velocity. 
The green/middle curve is the "null" profile. 
The Stokes V and null curves are shifted vertically and expanded by a factor of 
500. Dashed lines indicate the continuum level for Stokes I, and the zero level for the circular and null polarization.}
\label{fig:stokesv}
\end{figure}

For each spectrum, both Stokes I \& V parameters were processed using the LSD cross-correlation method  \citep{Do97}.
Using a line mask computed from a stellar atmospheric model with T$_{\rm eff}$=10,000~K and $\log g$ = 4.0 \citep{Kur93}, we
calculated LSD line profiles from a total of 1,200 photospheric lines. The multiplex gain in the S/N from the raw spectra to the
LSD mean profiles is about 30, reducing the noise level of the cross-correlation profiles to between $\sigma = 3$ and $7\times 10^{-5}I_{c}$, where $\sigma$ is the standard deviation of the noise and  $I_{c}$ stands
for the intensity of continuum. Since no signature was observed above noise level in individual Stokes V LSD profiles, we then calculated an 
average of the 257 profiles, where each profile is weighted by the square of its S/N. In this global profile, the noise in the Stokes V 
parameter is further decreased to  $\sigma = 2\times 10^{-6}I_{c}$ (Fig. \ref{fig:stokesv}) and a signature is now observed in circular polarization 
with an amplitude of $10^{-5}I_{c}$ (that is $5$ times the noise level). Running a $\chi^2$ test on the signature \citep{Do92}, we found a reduced $\chi^2$ of 3.5, which corresponds
to a false-alarm probability of $3\times 10^{-11}$.

\begin{figure}
\centering
\includegraphics[width=8cm]{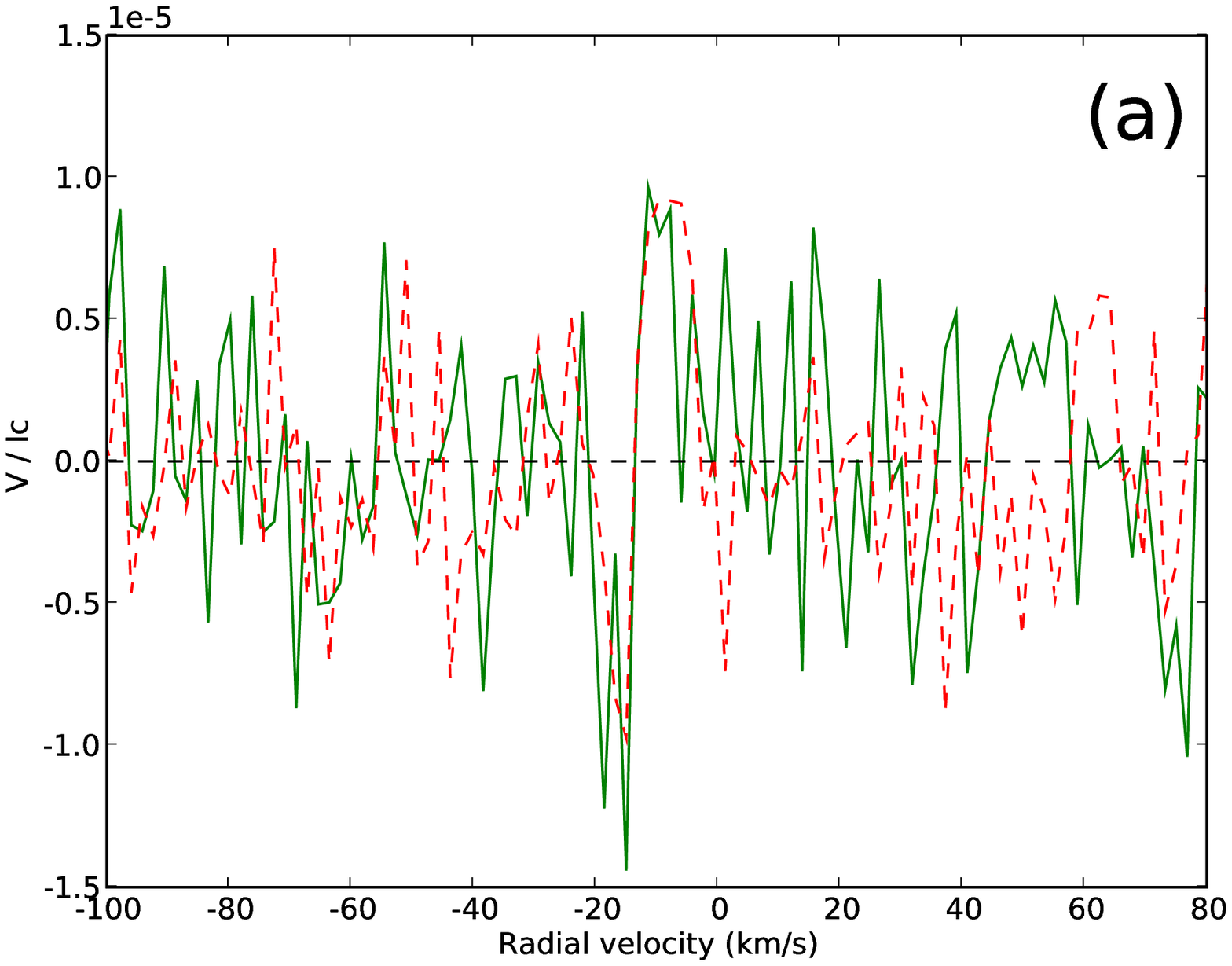}
\includegraphics[width=8cm]{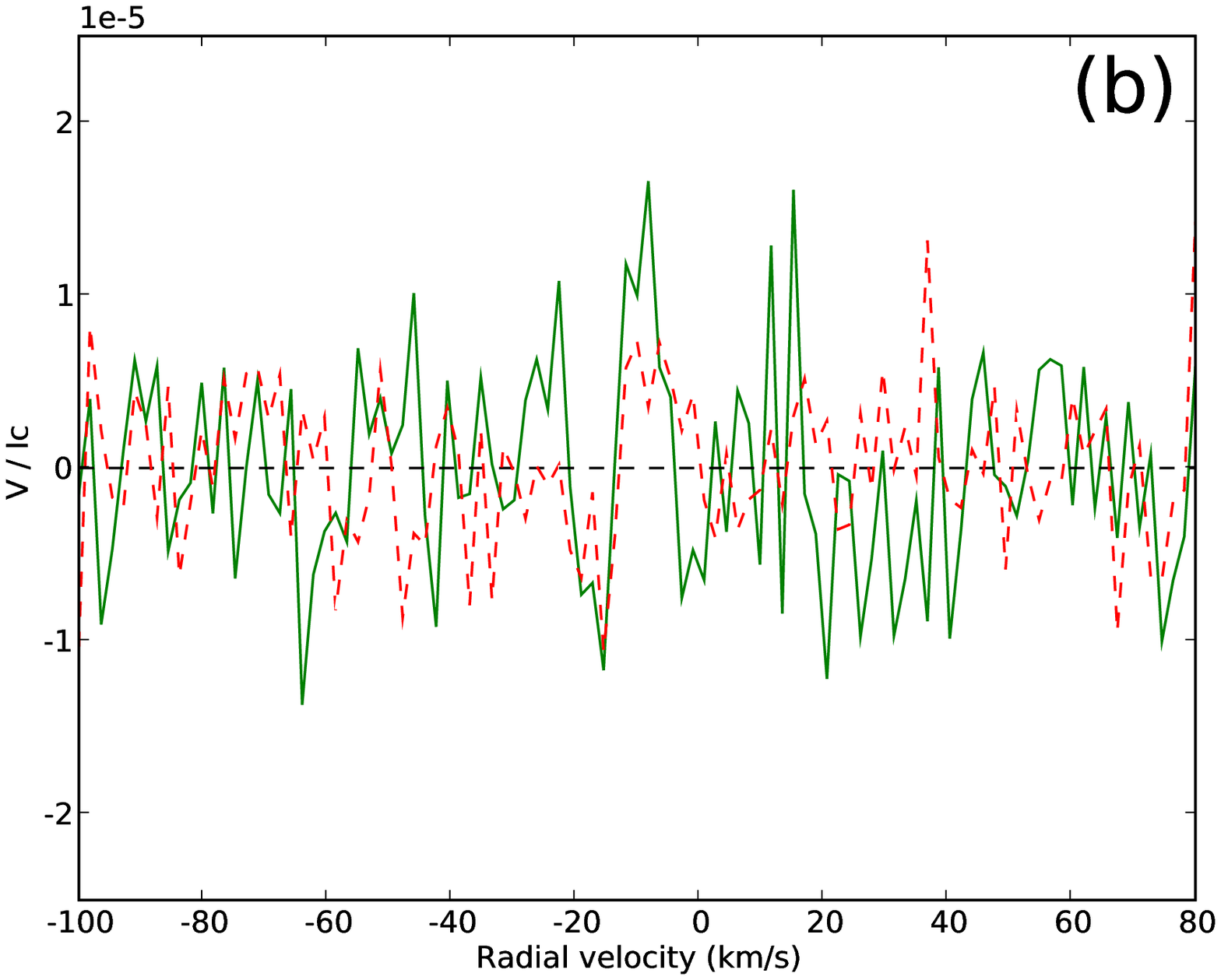}
\caption{(Color online) 
(a) Averaged LSD Stokes V line profiles obtained from two independent subsets of the whole observing material, 
containing respectively the first (green/continuous) and second (red/dashed) half of the observing run.
(b) Stokes V  
profiles obtained from spectral lines with Land\'e factors higher  
(resp. lower) than 1.2 (green/full and red/dashed lines, respectively).}
\label{fig:2epochs}
\end{figure}

Various tests have been performed to make sure that the polarization
is of stellar origin and not due to an artefact of the instrument or the reduction process.
This is particularly important in the present case, as the amplitude of the polarized signal is the lowest
detected by NARVAL to date.
A strong test to discard the possibility of a spurious signal is the "null" profile
calculated from a different combination of the
four sub-exposures constituting the polarimetric sequence \citep{Do97}. As shown in Fig.  \ref{fig:stokesv},
no detectable counterpart of the Stokes V signal is seen in the "null" profile (note that a similar conclusion is reached
by calculating another null
profile (not shown here) from another possible combination of the sub-exposures).
We then checked that the signal possesses the expected properties
of a stellar polarized signal.
First, we split the whole time-series into two independent subsets,
containing respectively the first and second half of the observing run, both subsets having equivalent signal-to-noise ratios.
As can be seen in Fig. \ref{fig:2epochs}a, the polarized signal is present in both sets, the false-alarm
probabilities based on the $\chi^2$ test being $5\times 10^{-6}$ and $6\times 10^{-3}$, respectively.
Second, we built two line-lists from the atmospheric model, containing all spectral
lines with Land\'e factors respectively higher and lower than $g_c = 1.2$. The
Stokes V profiles computed from the two line-lists are plotted in Fig. \ref{fig:2epochs}b.
As expected, the amplitude of the polarized signal appears
higher for the high Land\'e factor lines than for the low Land\'e factor lines.
The peak-to-peak amplitudes of the polarized signals taken inside the line profile are respectively $2.8
\times 10^{-5}I_c$ and $2.0\times 10^{-5}I_c$. Their difference slightly exceeds the
noise level ($\sigma = 6\times 10^{-6}I_c$ and $5\times 10^{-6}I_c$, respectively) and
their ratio is roughly consistent with the ratio of the average
Land\'e factors of the line-lists ($g_m=1.51$ and $g_m=0.94$ respectively), taking into account that the
corresponding Stokes I LSD profiles have a similar depth, within 10\%.
Third, we
checked that the signal was still consistently recovered when other
ways of splitting our line-list were considered (low versus high
excitation potential or low versus high wavelengths).
Finally, we tested the effect of changing the line mask. Indeed,
spectroscopic and interferometric studies of Vega have shown that
its surface temperature is inhomogeneous, due to the gravity darkening effect induced by its rapid rotation. According to
the latest model based on interferometric results \citep{Yo08}, the equatorial velocity of Vega is 274~km\,s$^{-1}$ and
the effective temperature and the gravity decrease from 9988~K and
$\log g = 4.07$ at the pole to 7600~K and $\log g = 3.5$ at the equator.
There is a significant discrepancy with the spectroscopic analysis of \citet{Ta08}, who find that the polar to equator temperature
difference is only $\sim\!900$~K while
the equatorial velocity is reduced to 175~km\,s$^{-1}$.
Taking the extreme case of a T$_{\rm eff}$=7500~K and $\log g$ = 3.5 stellar atmosphere,
we computed the LSD line profiles (not shown here) for
the associated line mask and this time obtained a false-alarm
probability of $10^{-1}$. The polarized signal is consistent with the one
derived from our previous (hotter) atmospheric model but the low significance of the detection
is due to the inadequacy of the line mask that results in
a
much higher noise level.

These tests strongly support that the signal is of stellar origin and therefore that
Vega possesses a magnetic field. 

The circularly polarized signal has the typical anti-symmetric shape of a Zeeman signature (Fig.  \ref{fig:stokesv}).
However, as compared to the width of the Stokes I line profile, it only shows up within 
a limited range of radial velocities about the line-center.
This suggests that the magnetic field distribution 
is axisymmetric and confined in the polar region. However, a more detailed analysis will be needed to specify the
surface field distribution of Vega. First, as the 257 spectra at our disposal cover a range of rotation
phases, Zeeman signatures from non-axisymmetric magnetic features, if any, are mostly averaged out from the time-averaged line profile.
Second, due to  Vega's temperature inhomogeneities, the weak line profiles range from 
flat-bottomed to "V" shapes \citep{Yo08,Ta08}.
As the LSD profile is obtained by assuming that all lines have a common profile,
its interpretation in terms of the surface field distribution is not straightforward in the present context.

We use the center-of-gravity method \citep{Ree79} to estimate the longitudinal magnetic field $B_l$ :
\begin{equation}
B_l = -2.14\times10^{11}\frac{\int vV(v)dv}{\lambda_0 g_m c\int(I_c - I(v))dv}
\end{equation}
\noindent where $v$ (km\,s$^{-1}$) is the radial velocity, $\lambda_0$ (nm) the mean wavelength of the line-list used to compute the LSD profiles,
$g_m$ the mean Land\'e factor and $c$ (km\,s$^{-1}$) the light velocity. The integration limits cover a $\pm30$~km\,s$^{-1}$ 
velocity range around the line centroid. Using this equation, we obtain $B_l = -0.6 \pm 0.3$~G. 

\section{Discussion}

Three basic features distinguish the present detection from previous measurements of magnetic fields in main-sequence stars of intermediate mass :
(i) It is the first time a magnetic field is detected in an A-type star which is not an Ap/Bp chemically peculiar star (if 
we exclude the debated field detections in a few Am and HgMn stars \citep{Ma93,Ma95} discussed in \citet{Shor02}).
(ii) The longitudinal magnetic field of Vega is smaller by about two orders of magnitude than 
the field of the most weakly magnetic Ap/Bp stars.
Indeed, the longitudinal field of a 300~G dipolar field aligned with the stellar rotation axis and viewed pole-on
is close to 100~G, that is about two orders of magnitude larger than the $0.6$~G field of Vega.
The longitudinal component of a dipolar field actually depends on its angle with respect to the rotation axis. But, whatever 
this angle, the amplitude of the circular polarization in the LSD Stokes V profile of a 300~G dipolar field 
will be more
than one order of magnitude larger than that of Vega.
(iii) The LSD Stokes V profile of Vega is also qualitatively distinct from LSD Stokes V profiles of Ap/Bp stars since
the polarized signal of Vega is concentrated in the weakly Doppler shifted regions of the projected stellar disk. 

These marked observational differences between Vega and the Ap/Bp magnetic stars suggest that we should consider Vega 
as a new type of magnetic A-type star.
As there is no reason to believe Vega is unique among A-type stars, Vega should be considered as the first member of
a new class of magnetic A-type stars.

The existence of such a new class might help understand some otherwise puzzling observations
of the pre-main-sequence and post-main-sequence intermediate mass stars.
The Herbig Ae/Be stars show a strong activity \citep[e.g.][]{Boh95} which has led investigators to suspect a widespread presence of magnetic fields in these stars \citep{Cat89}.
Nevertheless, these magnetic fields
have not been found, since only a small fraction of Herbig Ae/Be stars appears to host one \citep{Wade09}.
Note that a similar discrepancy between widespread activity and a small fraction of detected fields exists in OB stars \citep{Hen05,Schnerr08}.
A new class of magnetic A-type stars would shed new light on this issue. Indeed, the progenitors of these magnetic A-type stars
could be the Herbig Ae/Be stars where magnetic fields have not been detected yet, this non-detections being 
compatible with the fact that magnetic fields of the same intensity are much more difficult to detect in the faint Herbig Ae/Be stars 
than in a bright A-type star like Vega.
On the post-main-sequence side, the study of the white dwarf magnetic fields suggests that Ap/Bp stars are not sufficient to be the 
progenitors of magnetic white dwarfs \citep{Wick05}. A new class of magnetic A-type stars might also help to resolve this issue.

The consequences for Vega itself should also be considered. Its
magnetic field could indeed trigger active phenomena in its atmosphere.
Signs of spectroscopic variability have been reported, but have not been confirmed since \citep{Char85}.
On the other hand, despite its status as a photometric standard, a photometric variability of
$1-2 \%$ with occasional excursions to $4\%$ has been reported \citep[][]{Gr07}.
This might be produced  by photospheric temperature inhomogeneities induced by its magnetic field.
However, because of the near pole-on configuration of Vega, the variability would rather be due to intrinsic changes of the magnetic field
than to rotational modulation.

The origin of Vega's magnetic field could be attributed to one of the 
three mechanisms generally invoked for early-type stars, namely (i) the fossil field hypothesis, (ii) the envelope dynamo,
(iii) the convective core dynamo.
Let us first
consider the fossil field hypothesis, whereby the ISM magnetic field is confined
and amplified during stellar formation. It is regarded as the most consistent explanation of the magnetic fields 
observed in Ap/Bp stars \citep{Mo01}, but,
as proposed by
\citet{Au07}, it could also account for another population of stars hosting weak longitudinal magnetic fields.
Their argument is based on the fact that large-scale, organized magnetic field configurations are subjected
to a pinch-type instability driven by differential rotation \citep{Tay73,Sp99} when the magnetic field drops below a critical value. 
Consequently, for a distribution of large scale organized fields of different strengths issued from the star formation process,
the instability would produce
a magnetic dichotomy between
a population of strong and stable large scale fields like in Ap/Bp stars and another population where the destabilized
configuration is now structured at small length scales, thus resulting in a weak longitudinal field.
A simple estimate of the critical field has been found to be consistent with the reported lower limit of Ap/Bp stars. 
Here, both the detection of a very small longitudinal field in Vega and the gap between this field and the lowest magnetic fields of Ap/Bp stars 
reinforce this scenario.
Nevertheless, this scenario is not complete as it does not say what happens
to the destabilized field configuration, which could either decay or be regenerated by a dynamo.

The magnetic field of Vega could indeed be generated by an envelope dynamo where the energy source is the rotation of the star. 
Following \citet{Sp02}, the dynamo loop initiated by the differential rotation could be closed by the pinch-type instability mentioned
above.
This interesting possibility has been investigated by numerical simulations
in a simplified cylindrical configuration \citep{Brai06} and in a solar context \citep{Za07}, leading to opposite outcomes.
Simulations in more realistic conditions for A-type stars are clearly needed to test this envelope dynamo.
An important issue concerns the origin of the envelope differential rotation, which is a basic ingredient of this dynamo
but which is not forced by a strong stellar wind in A-type stars, 
contrary to what is expected to occur in
OB and Herbig Ae/Be stars \citep{Lign96}.
The third possibility is a dynamo in the convection core. While magnetic fields are likely to be generated there,
an efficient mechanism 
to transport it throughout the radiative envelope to the star surface has not yet clearly been identified \citep{Mac04}.
We note that in the three cases considered, the magnetic field is expected to be structured at small
scales and also probably variable in time.
This calls for a spectropolarimetric monitoring of Vega that will
investigate
the surface distribution and the temporal variation of its magnetic field.

\section{Conclusion}

A circularly polarized signal has been detected by accumulating a large number of high-quality echelle spectra of Vega
with the NARVAL spectropolarimeter. The data analysis strongly supports a stellar origin
of the polarization and thus the presence of a magnetic field on Vega.
Due to the unprecedented low level of the detected polarization,
new independent measurements will still be important to confirm this result.
A magnetic field on Vega suggests that other A-type stars which are not Ap/Bp stars host weak magnetic fields and that
their study can shed a new light on early-type star magnetism.
While a spectropolarimetric survey of bright A-type stars will be necessary to find these stars,
a detailed investigation
of Vega's magnetic field should also provide clues to the origin of this magnetism.

\begin{acknowledgements}
We thank the team of the Telescope Bernard Lyot for providing service observing and the referee, 
Prof. J. Landstreet, for useful comments which helped to improve
the manuscript.
\end{acknowledgements}

\end{document}